\documentclass[conference]{IEEEtran}

\usepackage{xcolor}

\usepackage{acronym}

\usepackage{graphicx}

\usepackage[capitalise,noabbrev]{cleveref}

\usepackage[font=footnotesize]{caption}
\usepackage{subcaption}

\usepackage[load-configurations=binary,detect-all,binary-units=true,range-phrase=--,per-mode=symbol]{siunitx}
\acrodef{3GPP}{Third Generation Partnership Project}
\acrodef{5G}{Fifth Generation}
\acrodef{NR}{New Radio}
\acrodef{eMBB}{enhanced Mobile Broadband}
\acrodef{URLLC}{Ultra Reliable Low Latency Communications}
\acrodef{mMTC}{massive Machine Type Communications}
\acrodef{ITU-R}{International Telecommunication Union, Radiocommunication Sector}
\acrodef{WP5D}{Working Party 5D}
\acrodef{IMT-2020}{International Mobile Telecommunications 2020}
\acrodef{LTE-M}{Long Term Evolution for Machines}
\acrodef{NB-IoT}{Narrowband Internet of Things}
\acrodef{ETSI}{European Telecommunications Standards Institute}
\acrodef{TC}{Technical Committee}
\acrodef{DECT}{Digital Enhanced Cordless Telecommunications}
\acrodef{FT}{Fixed Termination}
\acrodef{PT}{Portable Termination}
\acrodef{RD}{Radio Device}
\acrodef{CP}{Cyclic Prefix}
\acrodef{FDMA}{Frequency Division Multiple Access}
\acrodef{TDMA}{Time Division Multiple Access}
\acrodef{TDD}{Time Division Duplex}
\acrodef{CRC}{Cyclic Redundancy Check}
\acrodef{RF}{Radio Frequency}
\acrodef{FFT}{Fast Fourier Transform}
\acrodef{STF}{Synchronization Training Field}
\acrodef{DF}{Data Field}
\acrodef{GI}{Guard Interval}
\acrodef{PCC}{Physical Control Channel}
\acrodef{PDC}{Physical Data Channel}
\acrodef{DRS}{Demodulation Reference Signal}

\acrodef{OFDM}{Orthogonal Frequency-Division Multiplexing}
\acrodef{PDU}{Protocol Data Unit}
\acrodef{HARQ}{Hybrid Automatic Repeat Request}
\acrodef{RIT}{Radio Interface Technology}
\acrodef{TBS}{Transport Block Size}
\acrodef{LOS}{Line-of-Sight}
\acrodef{NLOS}{Non Line-of-Sight}
\acrodef{UE}{User Equipment}
\acrodef{PDP}{Power Delay Profile}
\acrodef{DS}{Doppler Spread}
\acrodef{MRC}{Maximum Ratio Combining}
\acrodef{BER}{Bit Error Rate}
\acrodef{PER}{Packet Error Rate}
\acrodef{SNR}{Signal-to-Noise Ratio}
\acrodef{LTE}{Long Term Evolution}
\acrodef{D2D}{Device-to-Device}
\acrodef{SISO}{Single Input Single Output}
\acrodef{SIMO}{Single Input Multiple Output}
\acrodef{MIMO}{Multiple Input Multiple Output}
\acrodef{AWGN}{Additive White Gaussian Noise}
\acrodef{MCS}{Modulation and Coding Scheme}
\acrodef{QAM}{Quadrature Amplitude Modulation}
\acrodef{QPSK}{Quadrature Phase-Shift Keying}
\acrodef{BPSK}{Binary Phase-Shift Keying}

\newcommand{\Matlab}{\textsc{Matlab}}

\begin{document}

\title{Link-Level Performance Evaluation of IMT-2020 Candidate Technology: DECT-2020 New Radio}

\author{
	\IEEEauthorblockN{Maxim Penner, Muhammad Nabeel, and J{\"u}rgen Peissig}
	\IEEEauthorblockA{Institute of Communications Technology, Leibniz Universit{\"a}t Hannover, Germany\\
      \texttt{\{penner,nabeel,peissig\}@ikt.uni-hannover.de}}
    }%
    
\begin{table*}
``This work has been submitted to the IEEE for possible publication. Copyright may be transferred without notice, after which this version may no longer be accessible.''
\end{table*}

\maketitle

\begin{abstract}
The ETSI has recently introduced the DECT-2020 New Radio (NR) as an IMT-2020 candidate technology for the mMTC and URLLC use cases.
To consider DECT-2020 NR as an IMT-2020 technology, the ITU-R has determined different independent evaluation groups to assess its performance against the IMT-2020 requirements.
These independent evaluation groups are now in process of investigating the DECT-2020 NR.
In order to successfully assess a technology, one important aspect is to fully understand the underlying physical layer and its performance in different environments.
Therefore, in this paper, we focus on the physical layer of DECT-2020 NR and investigate its link-level performance with standard channel models provided by the ITU-R for evaluation.
We perform extensive simulations to analyze the performance of DECT-2020 NR for both URLLC and mMTC use cases.
The results presented in this work are beneficial for the independent evaluation groups and researchers as these results can help calibrating their physical layer performance curves.
These results can also be used directly for future system-level evaluations of DECT-2020 NR.
\end{abstract}

\acresetall

%

\section{Introduction}
\label{sec:intro}

Due to growing number of new applications day by day and their advanced requirements, the mobile communication industry witnesses a generation transition almost after every decade.
To process this transition smoothly, different organizations regularly work together to create new standards, technical specifications, and regulations~\cite{dahlman20205g}.
In the last years, the \ac{ITU-R} \ac{WP5D} has set requirements for \ac{IMT-2020} to support the \ac{5G} use cases involving \ac{URLLC}, \ac{mMTC}, and \ac{eMBB}~\cite{ITU_2410}.
By targeting one or more of these \ac{5G} use cases, \ac{3GPP} has developed technologies such as \ac{NR}, \ac{NB-IoT}, and \ac{LTE-M} that meet the requirements set by \ac{ITU-R}~\cite{3GPP_self,ITU_2150}.

One of the candidate technologies currently being evaluated, submitted by \ac{ETSI} \ac{TC} \ac{DECT} and \ac{DECT} Forum to \ac{ITU-R}, is \ac{DECT}-2020 \ac{NR}~\cite{ETSI_636_1}.
The {DECT}-2020 \ac{NR} is a \ac{RIT} that mainly focuses on applications that fall between the \ac{URLLC} and \ac{mMTC} extremes, e.g., presence monitoring.
The \ac{DECT}-2020 \ac{NR} offers local deployment options without the need of a separate network infrastructure or network planning, and supports autonomous and automatic operation once deployed.
Hence, minimal deployment or maintenance effort is required making it an attractive technology.
The legacy \ac{DECT} standard was originally designed to support cordless telephone systems in early 1990s, but is continually updated in response to the developments in the technology.
Its latest development, {DECT}-2020 \ac{NR}, targets local area wireless applications, and offers star and mesh network topology to support \ac{URLLC} and \ac{mMTC} use cases, respectively.
The {DECT}-2020 \ac{NR} is operated in frequency bands below \SI{6}{\giga\hertz}, and despite its advanced physical layer numerology and medium access control algorithms, it can coexist with the legacy \ac{DECT} in current frequency bands allocated to legacy \ac{DECT}.

The \ac{ETSI} \ac{TC} \ac{DECT} has already provided detailed technical specifications of \ac{DECT}-2020 \ac{NR}~\cite{ETSI_636_3}.
However, to include \ac{DECT}-2020 \ac{NR} as an \ac{IMT-2020} technology, \ac{ITU-R} has determined different independent evaluation groups to assess its performance against the \ac{IMT-2020} requirements.
These independent evaluation groups are now in process of investigating the \ac{DECT}-2020 \ac{NR}.

In order to successfully assess a technology, one important aspect is to fully understand the underlying physical layer and its performance in different environments.
Therefore, in this paper, we first explain the \ac{DECT}-2020 \ac{NR} in detail and then analyze the link-level performance under standard channel models.
We have been already collaborating with the proponents for the self-evaluation of \ac{DECT}-2020 \ac{NR}, and we believe that the results presented in this paper will be beneficial to independent evaluation groups as well as other researchers.
These results will help them in understanding the physical layer of \ac{DECT}-2020 \ac{NR} better, in calibrating their performance curves, and in analyzing system-level performance of \ac{DECT}-2020 \ac{NR} by directly considering these results in their evaluation for different considered channel conditions.

Our main contributions can be summarized as follows:
\begin{itemize}
    \item We present a comprehensive overview of the newly introduced \ac{DECT}-2020 \ac{NR} standard (\cref{sec:dect2020}).
    \item We implement its complete physical layer and perform extensive simulations to evaluate the performance with standard channel models provided by \ac{ITU-R} (\cref{sec:implementation}).
    \item Finally, we show the link-level performance of \ac{DECT}-2020 \ac{NR} for both \ac{URLLC} and \ac{mMTC} use cases, which serve as comparative values and can be used directly for future system-level evaluations (\cref{sec:results}).
\end{itemize}

%

\section{Background and Related Work}
\label{sec:background}

\ac{DECT}-2020 \ac{NR} is a relatively new technology and, hence, does not have a vast literature.
Details of the \ac{DECT}-2020 \ac{NR} are provided by \ac{ETSI} \ac{TC} \ac{DECT} in~\cite{ETSI_636_1}, whereas an overview is presented in~\cite{kovalchukov2021dect}.
The main focus of~\cite{kovalchukov2021dect} is on the system-level evaluation of \ac{DECT}-2020 \ac{NR} and that also for the \ac{mMTC} use case only.
Recently, an independent evaluation group has also assessed this technology and provided the insights briefly in~\cite{dhanwani2020assessment}.
Due to lack of some details, authors were unable to conclude whether \ac{DECT}-2020 \ac{NR} meets the requirement specified to become an \ac{IMT-2020} technology.

To be considered as an \ac{IMT-2020} technology, \ac{DECT}-2020 \ac{NR} has to fulfil the requirement of \SI{99.999}{\percent} reliability, user plane latency of less than \SI{1}{\milli\second}, and capability of supporting more than one million devices per \si{\kilo\metre\squared}~\cite{ITU_2410}.
These requirements are set by \ac{ITU-R} \ac{WP5D} to support different \ac{5G} use cases.
The \ac{ITU-R} is a regulatory body and its responsibility is to ensure efficient and interference free operations of different radio communication systems, whereas \ac{WP5D} works within the \ac{ITU-R} and oversees IMT systems (i.e., 3G onwards) in particular~\cite{dahlman20205g}.
Both \ac{3GPP} and \ac{ETSI} are standards developing organizations and are part of \ac{ITU-R} \ac{WP5D}.
The \ac{3GPP} has been actively involved in developing technical specifications for mobile communication for more than two decades and it has developed technologies such as \ac{NR}, \ac{NB-IoT}, and \ac{LTE-M} to be the part of \ac{IMT-2020}~\cite{3GPP_self,ITU_2150}.
Each of these technologies usually targets applications that are bounded to a particular use case.
However, there are some applications that fall between the \ac{URLLC} and \ac{mMTC} extremes, e.g., remote light control and presence monitoring.
The reliability and delay requirements in these applications are much stricter than the ones described by \ac{mMTC} but are more relaxed in comparison to \ac{URLLC}.
Addressing such applications, \ac{ETSI} \ac{TC} \ac{DECT} has recently introduced the \ac{DECT}-2020 \ac{NR} as a candidate technology for the \ac{IMT-2020}.

The \ac{DECT}-2020 \ac{NR} offers a completely new physical layer numerology, advanced medium access control algorithms, and supports star as well as mesh network topology~\cite{ETSI_636_1,ETSI_636_3}.
However, in order to be considered as an \ac{IMT-2020} technology, its performance needs to be evaluated against the requirements defined in \ac{IMT-2020}.
In the literature, link-level performance of new technologies is usually evaluated by using standard channel models~\cite{sattiraju2020link}.
\ac{IMT-2020} has already provided different channel models to evaluate the candidate technologies in~\cite{ITU_2412}.
Therefore, in this work, we develop the physical layer of \ac{DECT}-2020 \ac{NR} and analyze its link-level performance.
The next section provides an overview of the \ac{DECT}-2020 \ac{NR} technology.

%

\section{DECT-2020 New Radio}
\label{sec:dect2020}

According to \ac{DECT}-2020 \ac{NR} technical specifications~\cite{ETSI_636_1}, any \ac{RD} in the network that has the capability of transmission and reception can be operated in either \ac{FT} mode or \ac{PT} mode or both simultaneously.
The \acp{RD} are free to choose any operation mode depending upon the local requirement.
A \ac{RD} operating in \ac{FT} mode coordinates local resources, and provides information to other \acp{RD} on how to initiate a connection and communicate with it.
The \ac{RD} in \ac{FT} mode is also responsible for routing the data either directly from the connected \acp{RD} to an external internet connection or through another \ac{RD} in \ac{FT} mode that has the access to the external internet connection.
The \ac{RD} in \ac{PT} mode simply connects to the \ac{RD} in \ac{FT} mode for an indirect association to the external internet connection.

\begin{figure}
     \includegraphics[width=.48\textwidth]{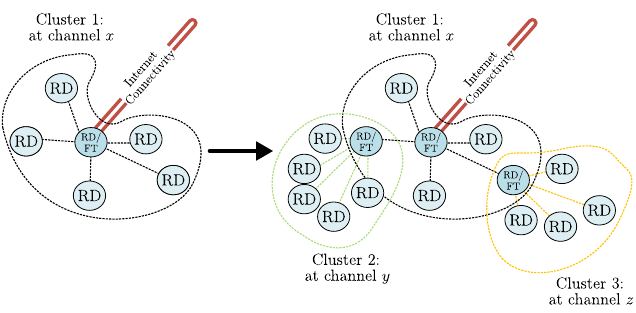}
     \caption{Formation of the mesh network topology.}
     \label{fig:mesh}
     \vspace{-2.5mm}
\end{figure}

A star topology network supports \ac{URLLC} use case and is formed by one \ac{RD} in \ac{FT} mode while other \acp{RD} directly connect to it in \ac{PT} mode.
Whereas, a mesh network topology is realized by adding more \acp{RD} that connect themselves to any of the network nodes, hence, offering a scalable solution for \ac{mMTC} use case as shown in \cref{fig:mesh}.
In the mesh network topology, all \acp{RD} are capable of routing data though they autonomously choose or change their role between routing and non-routing depending upon the local decisions.
Also, no central coordinator is needed and routing is based on a cost value rather than maintaining a routing table in each \ac{RD} resulting in autonomous routing.  
Moreover, there can be unrestricted number of \acp{RD} connected to any type or number of external internet connections in a single network offering more than one route for other \acp{RD} to choose, hence, minimizing the probability of outage.

In general, the \ac{DECT}-2020 \ac{NR} technology is targeted for frequency bands below \SI{6}{\giga\hertz}.
On physical layer, the system exploits \ac{CP} \ac{OFDM} in combination with \ac{TDMA} as well as \ac{FDMA} in a \ac{TDD} mode for communication~\cite{ETSI_636_3}.
The \ac{OFDM} subcarrier spacing is chosen between \SI{27}{\kilo\hertz}, \SI{54}{\kilo\hertz}, \SI{108}{\kilo\hertz}, or \SI{216}{\kilo\hertz} with a scaling factor $\mu =$ 1, 2, 4, or 8, respectively.
These different subcarrier spacings with a \ac{FFT} size $\beta =$ 64, 128, 256, 512, 768, or 1024 lead to a nominal \ac{RF} bandwidth between \SI{1.728}{\mega\hertz} and  \SI{221.184}{\mega\hertz}.

A single radio frame in \ac{DECT}-2020 \ac{NR} has a total duration of \SI{10}{\milli\second} and consists of 24 slots resulting in a slot duration of \SI{416.67}{\micro\second}.
Each of these slots can accommodate 10, 20, 40, or 80 \ac{OFDM} symbols depending upon the value of $\mu$.
A single slot is further divided into multiple subslots as discussed in~\cite{ETSI_636_3}.
Furthermore, on physical layer, the packet structure involves \ac{STF} and \ac{DF}, and is transmitted followed by a \ac{GI} to avoid overlapping of transmissions in consecutive time slots.
The \ac{STF} is composed of training data for time and frequency synchronization at the receiving side, whereas the \ac{DF} incorporates \ac{PCC}, \ac{PDC}, and \ac{DRS}.

The physical layer in \ac{DECT}-2020 \ac{NR} is mainly responsible for modulation and demodulation, error detection and correction, \ac{HARQ} soft combining, signal synchronization, and for providing data to upper layers.
Moreover, the physical layer also performs \ac{MIMO} antenna processing and offers a possibility of realizing transmit diversity and beamforming.
Error detection is performed by using a 16 or 24 bit \ac{CRC}, whereas for error correction, Turbo Coding is employed.
Finally, the supported modulations include \ac{BPSK}, \ac{QPSK}, as well as different \ac{QAM}, i.e., 16-\ac{QAM}, 64-\ac{QAM}, 256-\ac{QAM}, and 1024-\ac{QAM}.

%

\section{System Model}
\label{sec:implementation}

In \cite{ITU_2410}, the \ac{ITU-R} defines minimum requirements that radio interfaces must meet in each usage scenario of \ac{IMT-2020} as mentioned earlier.
For \ac{URLLC}, for instance, the minimum requirement for user plane latency is \SI{1}{\milli\second}, and the minimum requirement for reliability is a probability of success of \SI{99.999}{\percent} for layer 2 \ac{PDU} transmissions with a \ac{TBS} of at least 32 bytes.
Based on these constraints, suitable packet configurations of the \ac{DECT}-2020 \ac{NR} in \cite{ETSI_636_3} satisfying these latency and packet size requirements can be determined.
A selection of these configurations, which can be used for \ac{URLLC} and \ac{mMTC}, are listed in \cref{table_packet_formats}.
For the sake of simplicity, we call these packet configurations Format 0, Format 1, and Format 2.
All of these formats can be used for both \ac{mMTC} and \ac{URLLC} use cases, however, in this particular work, we use Format 0 for \ac{mMTC}, and Format 1 and Format 2 for \ac{URLLC} use case. 
The indicated number of \ac{HARQ} retransmissions is the maximum number with respect to the latency condition for the \ac{URLLC} use case.

\begin{table}[t]
    \renewcommand{\arraystretch}{1.3}
    \caption{\ac{DECT}-2020 \ac{NR} Packet Formats for \ac{URLLC} and \ac{mMTC}.}
    \label{table_packet_formats}
    \centering
    \begin{tabular}{c c c c}
        \hline
        \bfseries Packet Property & \bfseries Format 0 & \bfseries Format 1 & \bfseries Format 2\\
        \hline
        $(\mu,\beta)$       & (1,1) & (4,1) & (4,2)\\
        \ac{OFDM} symbols   & 10 & 10 & 10\\
        Transport Block Size                 & 296 & 368 & 288\\
        Modulation          & QPSK & QPSK & BPSK\\
        Code rate           & 1/2 & 3/4 & 1/2\\
        Bandwidth (\si{\mega\hertz})     & 1.728 & 6.912 & 13.824\\
        Subcarriers         & 64 & 64 & 128\\
        \ac{HARQ} retrans.  & 0 & 1 & 1\\
        Transmit antennas         & \multicolumn{3}{c}{1 or 2 (TX diversity)}\\
        \hline
    \end{tabular}
\end{table}

Furthermore, the \ac{ITU-R} also specifies how the suitability of candidate technologies for \ac{IMT-2020} can be tested on the physical layer~\cite{ITU_2412}.
For link-level evaluation, different channel models are provided, each consisting of path delays and corresponding average path gains for both \ac{NLOS} and \ac{LOS} scenarios.
The maximum Doppler frequency can be inferred from given carrier frequencies and the maximum velocity of \acp{UE}.
The Doppler shifts are then distributed according to the Jake's spectrum.
Thus, the combination of a \ac{PDP} and a \ac{DS} results in a doubly selective channel.
The channel parameters used in our simulations are summarized in \cref{table_channel_parameters}.
Because the time distances between individual taps of the provided \acp{PDP} in \cite{ITU_2412} are partly shorter than one sample, all \ac{OFDM} signals are oversampled to \SI{27.648}{\mega\hertz} regardless of the net bandwidth.

\begin{table}[t]
    \renewcommand{\arraystretch}{1.3}
    \caption{Channel Parameters of Doubly Selective Channel.}
    \label{table_channel_parameters}
    \centering
    \begin{tabular}{c c c}
        \hline
        \bfseries Channel Property & \bfseries Non Line-of-Sight & \bfseries Line-of-Sight\\
        \hline
        Power Delay Profile     & TDL-iii & TDL-v\\
        RMS Delay Spread        & \SI{363}{\nano\second} & \SI{93}{\nano\second}\\
        K factor                & - & \SI{9}{\decibel}\\
        Carrier Frequency       & \multicolumn{2}{c}{\SI{700}{\mega\hertz} and \SI{4}{\giga\hertz}}\\
        \ac{UE} Velocity        & \multicolumn{2}{c}{
        \SI{3}{\kilo\meter\per\hour} and \SI{30}{\kilo\meter\per\hour}}\\
        Doppler Spread          & \multicolumn{2}{c}{\SI{1.9}{\hertz} to \SI{111.2}{\hertz}}\\
        Simulation Bandwidth    & \multicolumn{2}{c}{\SI{27.648}{\mega\hertz}}\\
        Simulation Tool         & \multicolumn{2}{c}{\Matlab{} MIMOchannel}\\
        \hline
    \end{tabular}
\end{table}

\begin{table}[t]
    \renewcommand{\arraystretch}{1.3}
    \caption{Receiver Configuration.}
    \label{table_rx_config}
    \centering
    \begin{tabular}{c c}
        \hline
        \bfseries Receiver Property & \bfseries Receiver configuration\\
        \hline
        No. of Antennas             & 1, 2, or 4\\
        Diversity Combining         & Maximum Ratio Combining\\
        Time Synchronization        & on first channel tap\\
        Frequency Synchronization   & ideal\\
        Channel Estimation          & Wiener filter\\
        Channel Encoder/Decoder     & \Matlab{} \acs{LTE} Toolbox\\
        \hline
    \end{tabular}
\end{table}

The number of receive antennas that a base station or a \ac{UE} may utilize varies according to the use case.
In the case of \ac{mMTC}, base stations can have up to 64 antenna elements, whereas a single \ac{UE} can have a maximum of only 2 as described in~\cite{ITU_2412}.
For the \ac{URLLC} use case, a base station is allowed to have up to 256 antenna elements and a \ac{UE} up to 8.
These antennas can be used for beamforming, transmit or receive diversity, etc.
For our link-level evaluation, we consider only selected modes, i.e., a maximum of four receiver antennas and up to two transmitter antennas for diversity gain through space frequency block coding.
The complete configuration of the receiver is listed in \cref{table_rx_config}.

For packet synchronization, we assume that packets are received at the first channel tap.
This is a reasonable assumption as \ac{DECT}-2020 \ac{NR} exploits a slot-based system, therefore, time synchronization has a minor impact on the performance.
To test the design of the pilot pattern of \ac{DECT}-2020 \ac{NR}, we use a two-dimensional Wiener filter for channel estimation.
The channel coefficients are initially calculated by using an exponential \ac{PDP} with the largest conceivable delay spread and Jake's spectrum with the largest conceivable Doppler spread.
It is important to note that during an operation, the Wiener filter coefficients are fixed, even if the instantaneous channel statistics deviate from the aforementioned assumptions~\cite{hlawatsch2011wireless}.

The physical layer supports advanced channel coding (i.e., Turbo Coding) for both control and physical channels, and \ac{HARQ} with incremental redundancy.
The channel coding is based on \ac{LTE} with only minor differences, e.g., \ac{DECT}-2020 \ac{NR} can incorporate two maximum code block sizes instead of one as is the case with the \ac{LTE}~\cite{ETSI_636_3}.
The high similarity makes it possible to use existing encoders and decoders.
Therefore, we use the channel coding from the \ac{LTE} toolbox in \Matlab{} and incorporate it into our model in compliance with the \ac{DECT}-2020 \ac{NR} standard.

%

\section{Results and Discussion}
\label{sec:results}

To evaluate the performance of our system with the configurations presented in \cref{sec:implementation}, we conducted Monte Carlo experiments and measured the \ac{BER} and \ac{PER} at the receiver.
Each experiment ran until a convergence at a \ac{PER} of $\num{10e-5}$ or less was obtained, as this is the threshold that according to~\cite{ITU_2410} has to be reached for the \ac{URLLC} use case.

\begin{figure}[t]
	\centering
	\includegraphics[width=0.7\columnwidth]{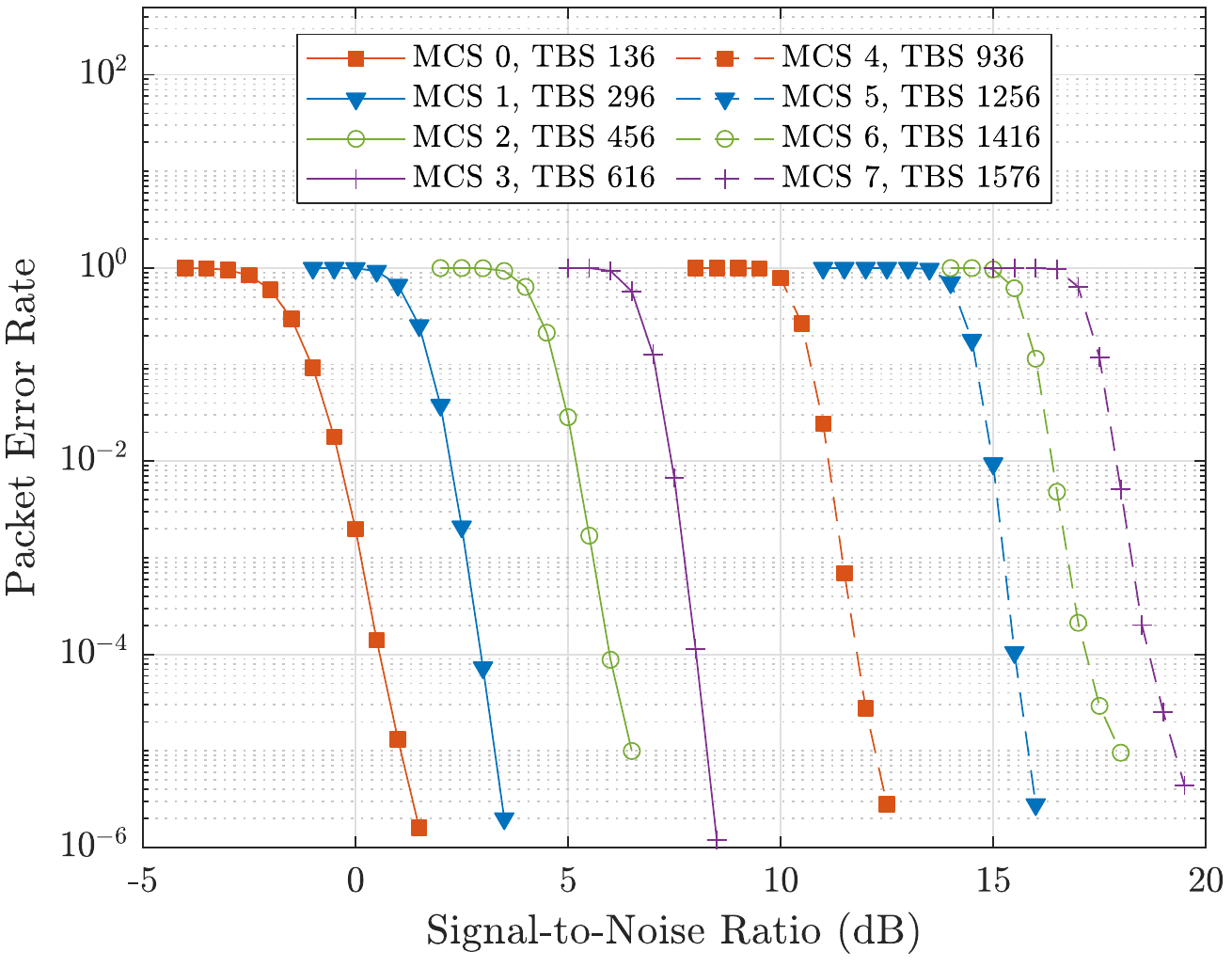}
	\caption{Packet error rate over different signal-to-noise ratio values in the AWGN channel under the assumption of perfect channel knowledge.}
	\label{fig:awgn}
\end{figure}

We first evaluate the performance of the first ten \acp{MCS} from~\cite{ETSI_636_3} over an \ac{AWGN} channel, and plot the results in \cref{fig:awgn}.
The \ac{MCS} 0 corresponds to \acs{BPSK} with code rate $1/2$, whereas \ac{MCS} 1 and \ac{MCS} 2 represent \ac{QPSK} with code rate $1/2$ and $2/3$, respectively.
Similarly, \ac{MCS} 7 corresponds to 64-\ac{QAM} with code rate $5/6$ and so on.
The size of the packet is fixed to $(\mu, \beta) = (1,1)$ and a single slot is considered, which implies that the increasing modulation order also increases the \ac{TBS}.
As expected, for higher modulation order, high \ac{SNR} value is required to achieve the same \ac{PER} performance.
These results are inline with the theory and the trend is similar to what we observe for \ac{LTE}.
However, these results cannot be directly compared with \ac{LTE} due to marginally different \acp{TBS} here.
The performance curves shown in \cref{fig:awgn} can be taken as a reference in the future to verify different implementations of the channel coding within the scope of \ac{DECT}-2020 \ac{NR}.

\begin{figure}[t]
	\centering
	\includegraphics[width=0.7\columnwidth]{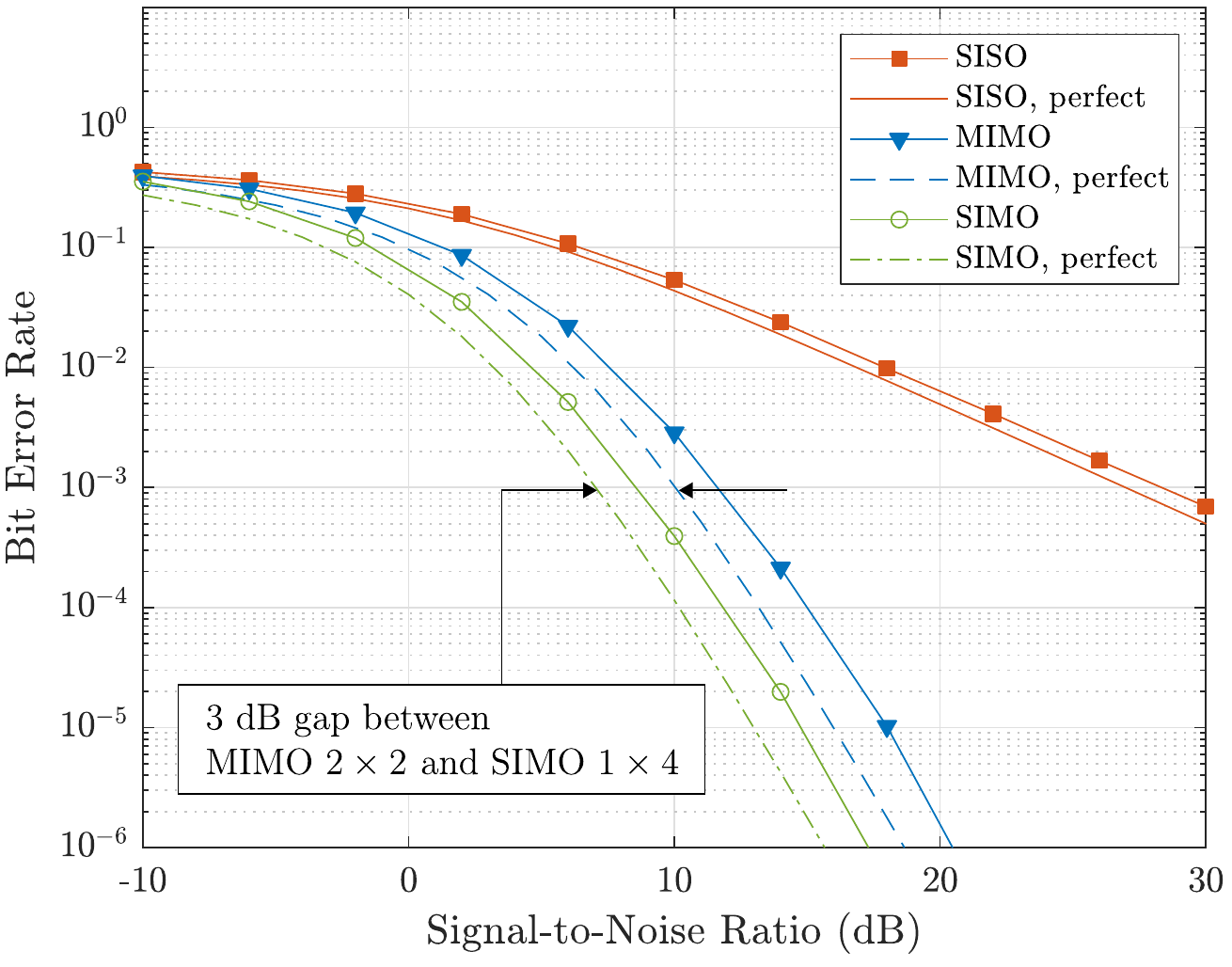}
	\caption{Bit error rate performance for \acs{QPSK} modulation in the Non Line-of-Sight with and without perfect channel knowledge.}
	\label{fig:comparison}
	\vspace{-4.5mm}
\end{figure}

As a next step, we analyze the performance of our Wiener filter used for channel estimation.
\Cref{fig:comparison} shows the actual measured \ac{BER} after using the Wiener filter as well as the \ac{BER} under the assumption of perfect channel knowledge.
For the latter, we assume a flat fading Rayleigh channel.
The results are shown for a \ac{SISO}, a $2\times 2$ \ac{MIMO}, and a $1\times 4$ \ac{SIMO} system.
For \ac{SISO}, the channel estimation of the Wiener filter makes the performance about \SI{1}{\decibel} worse in comparison to a system with perfect channel knowledge.
In the case of both \ac{MIMO} and \ac{SIMO}, this loss is around \SI{2}{\decibel}.
It is also interesting to note that there is a \SI{3}{\decibel} gap between \ac{MIMO} and \ac{SIMO} results.
This is because of the fact that the transmitter power must be split between the two transmit antennas in the case of \ac{MIMO}.
The comparison with closed-form solutions for the \ac{BER} in Rayleigh fading channels was used for all experiments, since this confirms the accuracy of the results as well as the suitability of the pilot pattern in \ac{DECT}-2020 \ac{NR}.

\begin{figure*}
	\centering
	\subfloat[Line-of-Sight (LOS)]{\label{fig:los}\includegraphics[width=0.7\columnwidth]{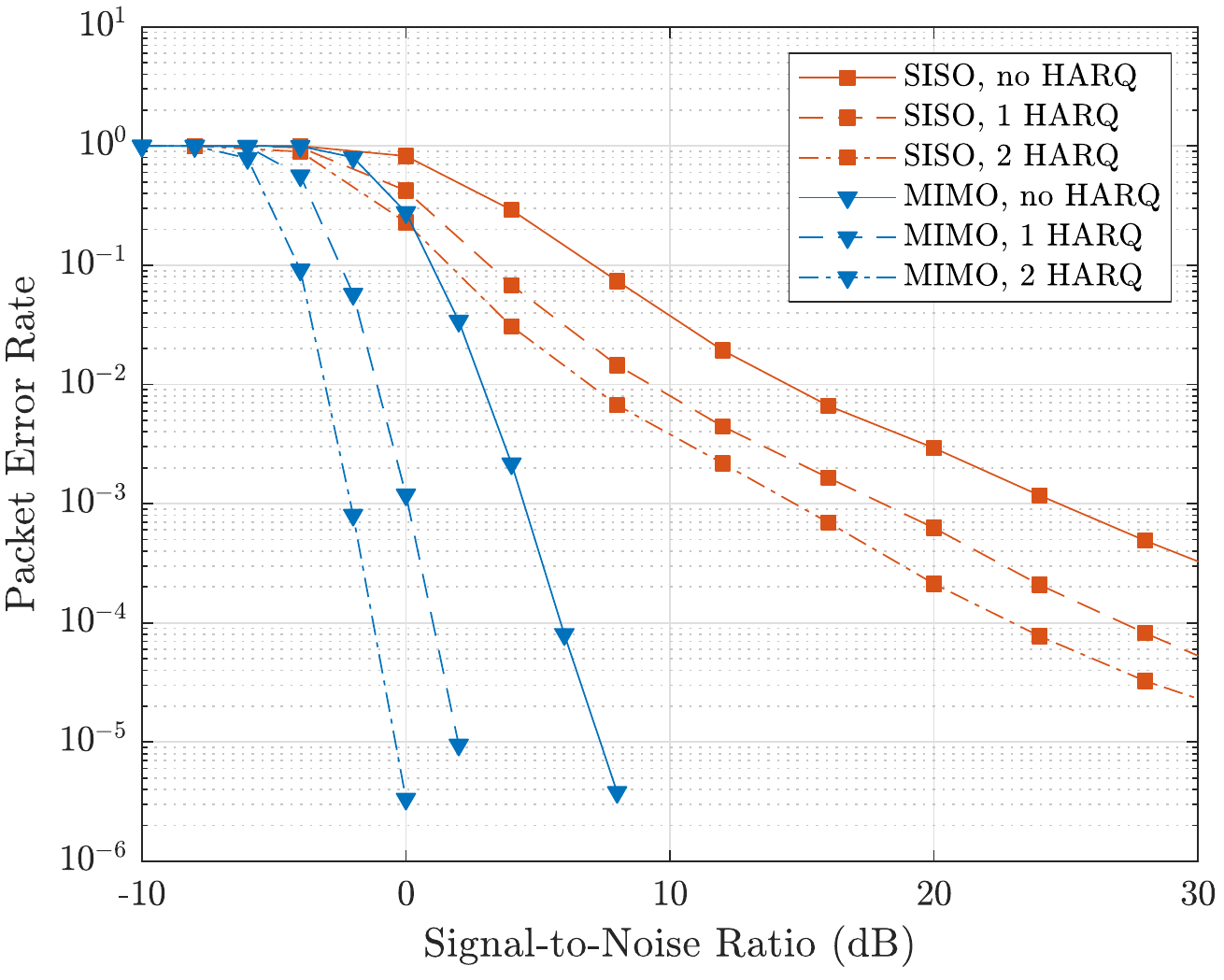}} \qquad \qquad
	\subfloat[Non Line-of-Sight (NLOS)]{\label{fig:nlos}\includegraphics[width=0.7\columnwidth]{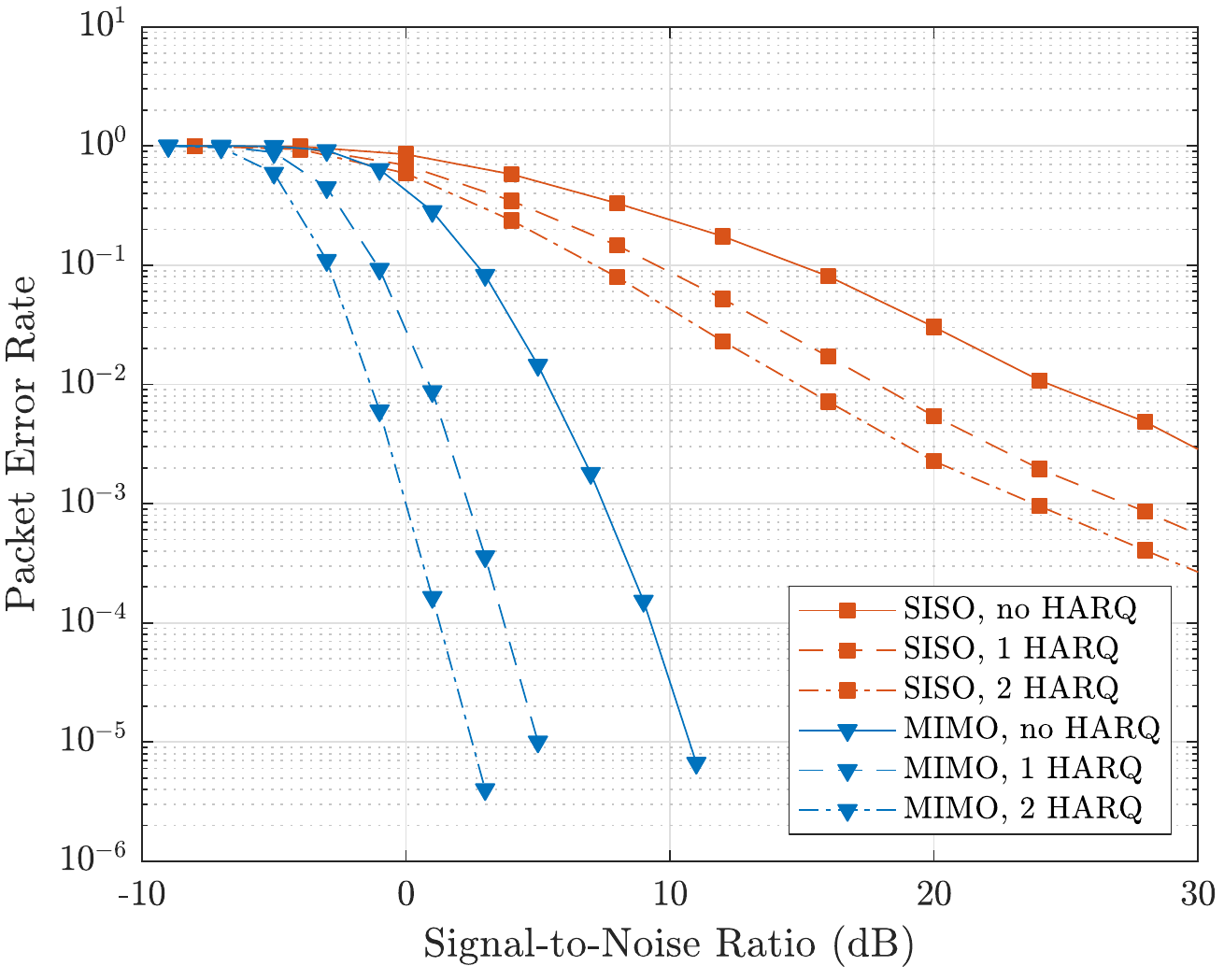}}
	\caption{Packet error rate for \acs{SISO} and $2\times 2$ \acs{MIMO} systems with Format 0. In the case of \acs{MIMO}, the system utilizes both transmit and receive diversity.}
	\label{fig:packet0}
\end{figure*}

We then investigate the performance of packet Format 0, i.e., the simplest among the considered formats, with both \ac{SISO} and $2\times 2$ \ac{MIMO} antenna configurations, and plot the results in \cref{fig:packet0}.
The results for \ac{LOS} and \ac{NLOS} scenarios are shown in \cref{fig:los} and \cref{fig:nlos}, respectively.
Within the context of \ac{DECT}-2020 \ac{NR}, a \ac{SISO} system could be deployed for \ac{D2D} communication within the mesh topology, where the devices must be as simple as possible in design.
As expected, the \ac{PER} in a \ac{SISO} case for both \ac{LOS} and \ac{NLOS} decreases only slowly since there is no diversity.
In the case of \ac{MIMO}, since diversity is employed at both ends of the transmission link, the \ac{PER} is highly improved.
For the \ac{NLOS}, we achieve a \ac{PER} of $\num{10e-5}$ without \ac{HARQ} at a \ac{SNR} of about \SI{11}{\decibel} and with two \ac{HARQ} retransmissions at a \ac{SNR} of about \SI{2.5}{\decibel}.

\begin{figure*}
	\centering
	\subfloat[Packet Format 1]{\label{fig:packet1}\includegraphics[width=0.7\columnwidth]{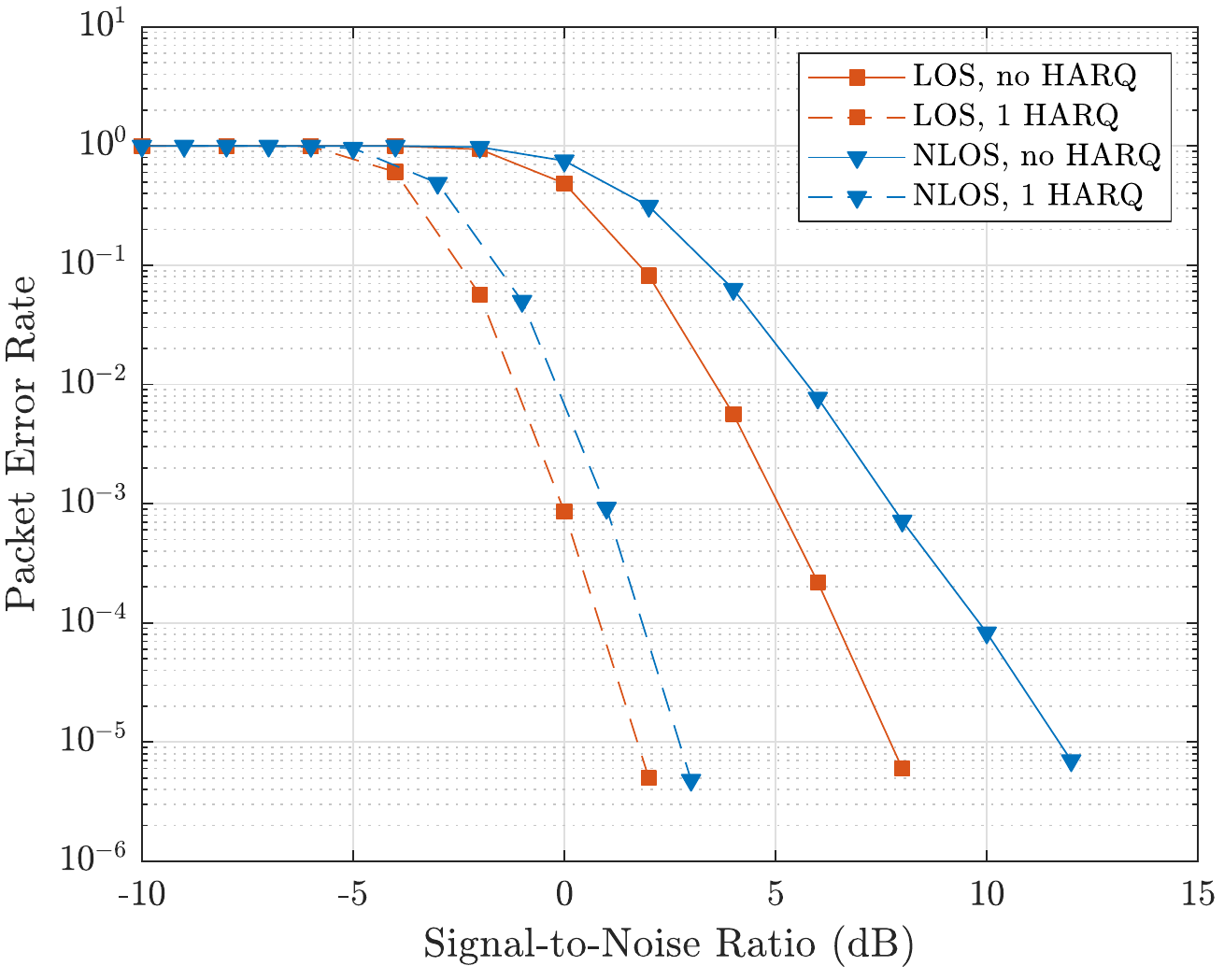}} \qquad \qquad
	\subfloat[Packet Format 2]{\label{fig:packet2}\includegraphics[width=0.7\columnwidth]{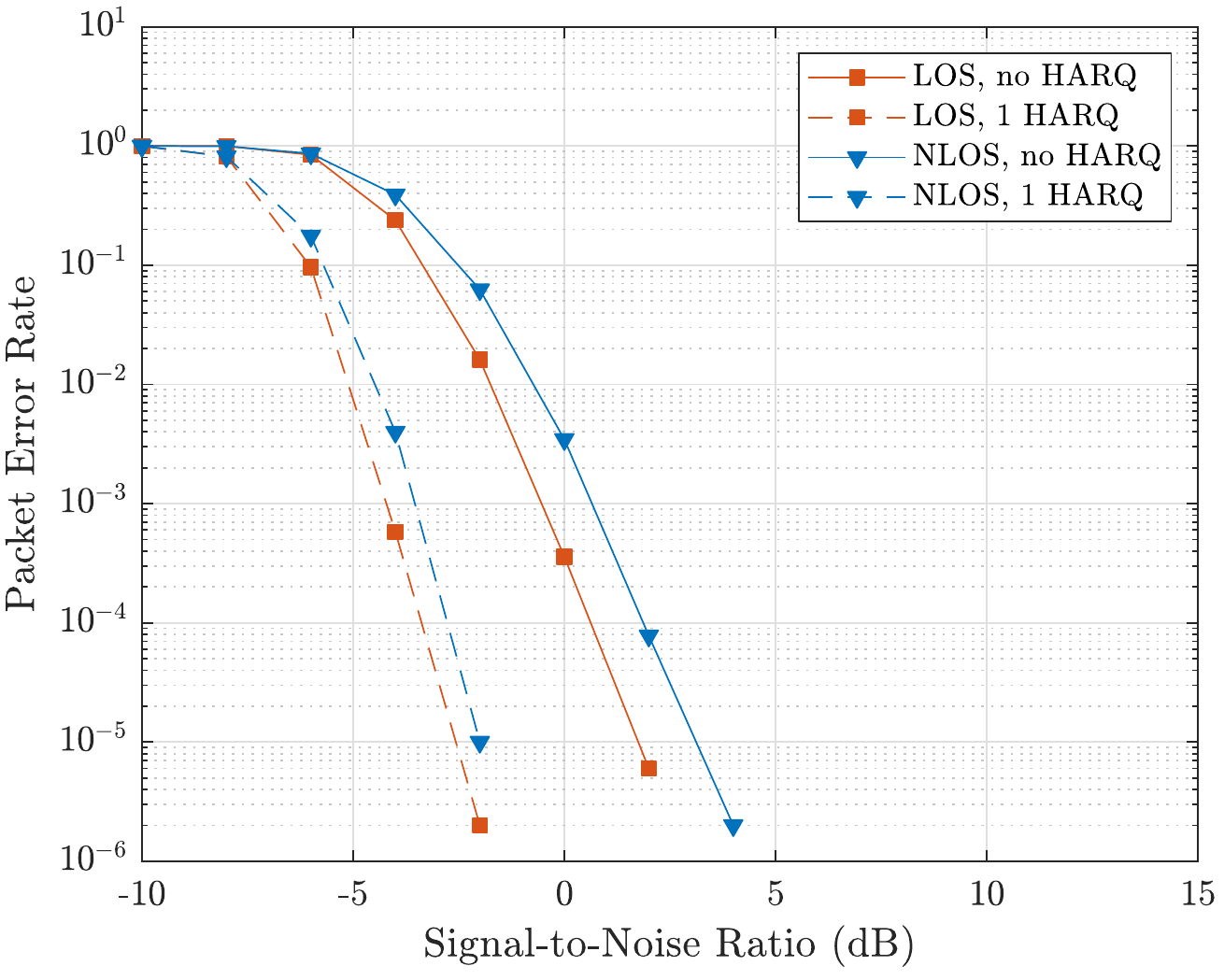}}
	\caption{Packet error rate for $1\times 4$ \ac{SIMO} system with different packet formats and under different channel conditions.}
	\label{fig:packet12}
	\vspace{-2.5mm}
\end{figure*}

Finally, we focus on \ac{URLLC} use case by considering \ac{SIMO} with one transmit and four receive antennas for maximum reliability, and present the results in \cref{fig:packet12}.
\Cref{fig:packet1} and \cref{fig:packet2} show the performance for packet Format 1 and Format 2, respectively, in \ac{LOS} and \ac{NLOS} scenarios.
It can be seen that in \cref{fig:packet2}, we achieve a \ac{PER} of $\num{10e-5}$ at a \ac{SNR} of \SI{3.5}{\decibel} for \ac{NLOS} and at \SI{2}{\decibel} for \ac{LOS} when using no \ac{HARQ}.
With one \ac{HARQ} retransmission we reach the same threshold at a \ac{SNR} of \SI{-2}{\decibel} and \SI{-2.5}{\decibel} for \ac{NLOS} and \ac{LOS}, respectively.
For Format 1 in \cref{fig:packet1}, the values are comparatively higher.
This is because we use \ac{QPSK} instead of \ac{BPSK} with half the signal bandwidth.

%

\section{Conclusion}
\label{sec:conclusion}

This work evaluated the link-level performance of the novel \ac{IMT-2020} candidate technology, \ac{DECT}-2020 \ac{NR}.
We first provided an overview of the \ac{DECT}-2020 \ac{NR} and presented 
possible packet configurations that meet latency and packet size requirements for \ac{URLLC} and \ac{mMTC} use cases.
We then considered doubly selective channel models as provided by the \ac{ITU-R} to evaluate the performance with different configurations of antenna, \ac{HARQ} retransmissions, and modulation and coding schemes.
Our experiments show that in all cases, receivers can be built that operate close to perfect channel knowledge.
This confirms that for the models tested, \ac{DECT}-2020 \ac{NR} is a well designed \ac{OFDM} system.
The results presented in this work can be used directly for future system-level evaluations by considering these \ac{PER} values over different \acp{SNR} as an input.

\bibliography{references}
\bibliographystyle{IEEEtran}

\end{document}